\documentclass[preprint,review,12pt]{elsarticle}
\usepackage{amsmath, amsthm, amssymb}
\usepackage{hyperref}
\usepackage{verbatim}
\usepackage{lineno}
\usepackage{epsfig,graphicx}
\usepackage{multirow}
\usepackage[english]{babel}

\makeatletter
\newcommand\gsl{\ifmmode\textsl{g}\else g\fi}
\newcommand{\const}{\operatorname{const}}
\newcommand{\sgn}{\operatorname{sign}}
\makeatother

\journal{Dynamics of Atmospheres and Oceans}

\begin{document}

\begin{frontmatter}

\title{Large-scale dynamics of equatorial thermal spots}

\author{V.~P.~Goncharov}
\ead{v.goncharov@rambler.ru}
\affiliation{A. M. Obukhov Institute of Atmospheric Physics RAS, 109017 Moscow, Russia}

\date{\today}

\begin{abstract}
For a model of a thin spherical layer of an incompressible rotating fluid subjected to the action of Coriolis and buoyancy forces, both analytical solutions in the form of stationary thermal spots moving along the equator and self-similar solutions are found. The influence of the parameters determining the shape of the spots on their propagation speed and direction is studied. It is found that the speed and direction of motion of the spots depend not only on the sign of their thermal contrast with the background flow, but also on the ratio of the semi-axes determining their shape. In particular, if the ratio of the equatorial semi-axis to the meridional one $b/a$ is equal to $\pi/2$, the spots are at rest. If the spots are elongated along the equatorial axis to such an extent that $b/a>\pi/2$, then ``cold'' spots move westward and ``hot'' spots move eastward. Under the condition $0.747\leq b/a<\pi/2$, the situation is reversed: ``hot'' spots move westward and ``cold'' spots move eastward. Self-similar solutions in the model are realized strictly in the form of circular thermal spots, and their radius varies according to a power law with a scaling exponent $k$, which is determined by the behavior of the cross-frontal buoyancy gradient on the spot contour. We note that the regime $k=1/4$ is realized under the assumption of a constant buoyancy gradient, while the regime $k=1/6$ corresponds to the conservation of total buoyancy.
\end{abstract}


\end{frontmatter}


\section{Introduction}\label{sec1}

The large-scale dynamics of thermal spots forming in a thin layer of an incompressible rotating fluid on a sphere or on the equatorial beta-plane under the influence of Coriolis and gravity forces can be studied within the framework of the approach known as contour dynamics~\cite{dg78,p92,z18}. The corresponding theoretical model developed in~\cite{g23} is based on the use of the variational principle of least action and perturbation theory with two small parameters $\lambda$, $\epsilon$, which control the smallness of temporal and spatial variations compared with the characteristic scales. As the latter, the half-period of rotation $T$ and the Rossby deformation radius $L$ are considered. It is shown that under the condition $3L\lambda/R=\epsilon^{4}$, where $\epsilon=l/L$, $R$ is the planet radius, and $l$ is the width of the frontal zone where the cross-frontal buoyancy gradient $\gamma$ is concentrated, the model admits the existence of thermal spots around which an intense jet flow is formed. In this case, perturbation theory in the first order turns out to be self-consistent and, after nondimensionalization, leads to the equations
\begin{equation}
X_{t}Y=\partial_{s}\frac{\gamma X_{ss}}{\left(X_{s}^{2}+Y_{s}^{2}\right)^{2}},\quad
Y_{t}Y=\partial_{s}\frac{\gamma Y_{ss}}{\left(X_{s}^{2}+Y_{s}^{2}\right)^{2}}.\label{eq:1}
\end{equation}
Here, the dynamical variables $X$ and $Y$, being functions of time $t$ and the parameter $s$, describe the contour of the thermal spot. Exactly the same equations hold in the equatorial beta-plane approximation~\cite{g23}. If in the planar model the coordinates $X$ and $Y$ are defined as
\begin{equation}
X=\epsilon x/L,\quad Y=\epsilon y/L,\label{eq:2}
\end{equation}
through the ordinary Eulerian coordinates $x$ and $y$ on the spot contour, then for a rotating spherical layer~\cite{g23}, as 
\begin{equation}
X=\epsilon\frac{R}{L}\vartheta,\quad Y=\epsilon\frac{R}{L}\sin\varphi,\label{eq:3}
\end{equation}
where $\vartheta$ and $\varphi$ are longitude and latitude. It should also be noted that, since the frontal zone around the spot, where the contour is located, is assumed to be sufficiently narrow, $l/L=\epsilon\ll1$, in the case when the frontal temperature jump $\Delta\theta$ (or density jump $\Delta\varrho$) does not vary along the contour, the definition of the cross-frontal buoyancy gradient, with the chosen method of scaling in the problem, reduces to the use of the rule
\begin{equation}
\gamma=-\sgn(\Delta\theta)=\sgn(\Delta\varrho).\label{eq:4}
\end{equation}

Let us note that in model \eqref{eq:1} there is no transition to a linear problem. The nonlinearity effect due to the beta-effect is singular in nature and does not vanish as $Y\rightarrow0$. Such nonlinearities are encountered in many natural media and metamaterials~\cite{r18}.

A comparative analysis of the equations obtained in~\cite{g21} for the planar model (without taking the beta-effect into account) and equations \eqref{eq:1} reveals not only the loss by the latter of the symmetry $Y\rightarrow-Y$ and the translational symmetry $Y\rightarrow Y+\const$, but also an additional (singular) nonlinearity due to the factor $Y$ on the left-hand sides of the equations. We note that the study of the nonlinear dynamics of equatorial thermal spots is of interest for many reasons, and primarily because, unlike waves, they are localized and possess a closed contour. Such a topology allows them to provide convective transport of heat and impurities along the equator in the atmosphere and ocean more efficiently than waves. It is quite possible that, like solitons in the context of nonlinear processes, equatorial thermal spots, on closer examination (in the distant future), may turn out to be structural elements of equatorial large-scale turbulence or support the mechanism of such climatic phenomena as El Ni\~no and La Ni\~na.

The structure of this article is as follows. In Section~\ref{sec2}, equations~\eqref{eq:1} are reformulated, with the help of a hodograph transformation and the use of signum functions, into a form convenient for constructing solutions with a closed contour. In Section~\ref{sec3}, steady-state solutions corresponding to thermal spots moving along the equator without changing shape at a constant speed are considered. Self-similar solutions for equatorial thermal spots, their shape, and possible scaling laws are discussed in Section~\ref{sec4}. In Section~\ref{sec5}, we summarize the results of the work.

\section{Hodograph transformation and the use of signum functions}\label{sec2}
 
The dependent variables $X$ and $Y$, which describe the contour dynamics of spots in the original model~\eqref{eq:1}, are functions of the parameter $s$ and time $t$. Let us consider the so-called hodograph transformation, which assigns $Y$ and $s$ as the new dependent variables and $X$ and $t$ as the independent variables. Then for the new equations in these variables we obtain
\begin{equation}
s_{t}Y=\gamma\partial_{X}\frac{s_{X}s_{XX}}{\left(1+Y_{X}^{2}\right)^{2}},\quad
Y_{t}Y=\gamma\partial_{X}\frac{s_{X}Y_{XX}}{\left(1+Y_{X}^{2}\right)^{2}}.\label{eq:5}
\end{equation}

\begin{figure}[ht!]
\centering{\includegraphics[width=\linewidth]{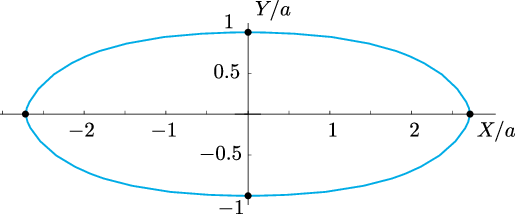}}
\caption{A thermal spot with a closed contour. The coordinate axes are normalized to the meridional semi-axis of the spot $a$. Turning points are highlighted, at which the derivative $Y_{X}$ change sign.}\label{fig1}
\end{figure}

Let us consider solutions with a closed contour that is mirror-symmetric with respect to the equatorial axis $X$ (see Fig.~\ref{fig1}). Since such solutions, being defined on a compact support, have turning points on the axis $X$, they are multivalued. One of the possibilities for formalizing such solutions is the use of signum functions.

In the general case, to define the signum function one can use the expression
\[
\sgn(G)=\left\{
\begin{array}{r}
1,\ if\ G>0;\\
0,\ if\ G=0;\\
-1,\ if\ G<0;%
\end{array}
\right.
\]
where the arguments of the sign function $G=G\left(X,Y\right)$ are chosen so as to ensure the correct alternation of sign when passing through the turning points.

In the context of our problem (taking into account that the turning points lie on the axis $X$), it is convenient to choose the function $\sgn(Y)$ as the signum function, and to consider the relation
\begin{equation}
s=X\sgn(Y)=\left\{
\begin{array}{r}
X,\ if\ Y>0;\\
0,\ if\ Y=0;\\
-X,\ if\ Y<0.%
\end{array}
\right.\label{eq:6}
\end{equation}
as the solution for the variable $s$. It is easy to see that, with this choice, on the one hand, the sign change is ensured when passing through the turning points on the axis at $Y=0$, and, on the other hand, the first of equations~\eqref{eq:5} is satisfied identically.

As a result, we arrive at a description with a single equation
\begin{equation}
|Y|Y_{t}=\gamma\partial_{X}\frac{Y_{XX}}{\left(1+Y_{X}^{2}\right)^{2}},\label{eq:7}
\end{equation}
in which $\gamma=1$ if ``cold'' spots are considered, $\gamma=-1$ if ``hot'' ones are considered, and to which boundary conditions must be adjoined that correspond to the model ideas about the behavior of the thermal spot contour at the turning points at $Y=0$.

\section{Steady-state solutions for equatorial spots.}\label{sec3} 

Let us consider solutions of equation \eqref{eq:7} defined on a compact support and corresponding to thermal spots moving along the equator without changing shape at a constant speed $c$. In this case, in the comoving coordinate system $(x,y)$, assuming that
\begin{equation}
X=ct+ax,\quad Y=ay(x),\label{eq:8}
\end{equation}
and using the meridional semi-axis of the thermal spot $a=\max|Y|$ as the scaling factor $a$, we obtain the ordinary differential equation
\begin{equation}
3p|y|y_{x}+\partial_{x}\frac{y_{xx}}{\left(1+y_{x}^{2}\right)^{2}}=0,\quad 
p=\frac{\gamma}{3}ca^{3}.\label{eq:9}
\end{equation}
describing the shape of the thermal spot.

Taking into account the mirror symmetry of the solutions of equation~\eqref{eq:9} with respect to the axes, it suffices to find its solution in the first quadrant of the coordinate plane, where $x\geq0$ and $1\geq y\geq0$. After integrating \eqref{eq:9} twice, we find
\begin{equation}
py^{3}-\frac{1}{1+y_{x}^{2}}+c_{1}y+c_{2}=0,\label{eq:10}
\end{equation}
with the boundary conditions
\begin{equation}
\left.y_{x}\right\vert_{y=1}=0,\quad
\left.y_{x}\right\vert_{y=0}=-\infty,\label{eq:11}
\end{equation}
Here the first of the conditions assumes that at some initial point, which without loss of generality can be taken as $x=0$, the function $y$ has an amplitude maximum $y=1$, and the second assumes that at the final point, where the function $y$, decreasing, vanishes, its derivative tends to minus infinity.

Conditions \eqref{eq:11} fix the integration constants:
\begin{equation}
c_{1}=1-p,\quad c_{2}=0,\label{eq:12}
\end{equation}
and thus equation \eqref{eq:10} is reduced to the form
\begin{equation}
py^{3}-\frac{1}{1+y_{x}^{2}}+y\left(1-p\right)=0,\label{eq:13}
\end{equation}
whence follows the integral representation for $x$:
\begin{equation}
x=-\int_{1}^{y}\sqrt{\frac{u\left(1-p\left(1-u^{2}\right)\right)}
{(1-u)\left(1+p\left( u+u^{2}\right)\right)}}du,\label{eq:14}
\end{equation}
where, in addition to the condition $1\geq y\geq0$, the inequality $-\frac{1}{2}<p\leq1$ holds, which in turn follows from the condition that the radicand in the integral is nonnegative.

The result of integrating \eqref{eq:14} can be represented analytically only for particular values of the parameter $p=0,1$. If $p=0$, the solution can be written in parametric form as
\begin{equation}
x=\varphi+\sin\varphi\cos\varphi,\quad
y=\cos^{2}\varphi,\label{eq:15}
\end{equation}
where $|\varphi|<\pi/2$ is a parameter. If $p=1$, the solution is expressed in terms of the Gauss hypergeometric function $\,_{2}F_{1}$:
\begin{multline}
x=-\int_{1}^{y}\frac{u^{3/2}}{\sqrt{1-u^{3}}}du=\\
=\sqrt{\pi}\frac{\Gamma\left(\frac{5}{6}\right)}{\Gamma\left(\frac{1}{3}\right)}-
\frac{2}{5}y^{5/2}\,_{2}F_{1}\left(\frac{1}{2},\frac{5}{6};
\frac{11}{6};y^{3}\right).\label{eq:16}
\end{multline}

\begin{figure}[t!]
\centering\includegraphics[width=\linewidth]{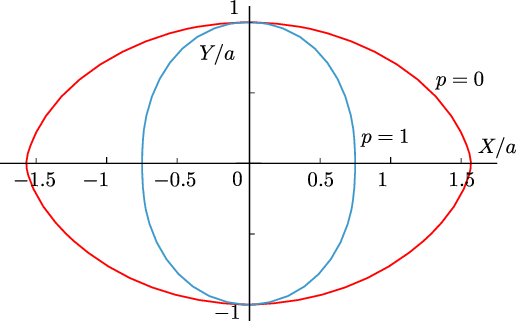}
\caption{Shape of thermal spots with the ``dispersion'' parameter $p=0,1$. The red contour corresponds to $p=0$, the blue one to $p=1$. The coordinate axes are normalized to the meridional semi-axis of the spot $a$.}\label{fig2}
\end{figure}
Taking into account the ``dispersion'' relation $p=\gamma ca^{3}/3$, it is easy to conclude (see Fig.~\ref{fig2}) that thermal spots of the first type are elongated equatorially and are at rest ($b/a=\pi/2$, $c=0$), while those of the second type, on the contrary, are elongated meridionally ($b/a=0.747$) and move with speed $c=3/a^{3}$ in the westward direction if they are ``hot'', and in the eastward direction if they are ``cold''.

In the general case, the deformation characteristic of a thermal spot can obviously be the quantity $D=b/a$ --- the ratio of the equatorial semi-axis $b$ to the meridional one $a$. On the basis of \eqref{eq:14} it is easy to find that
\begin{equation}
D(p)=\int_{0}^{1}\sqrt{\frac{u\left(1-p\left(1-u^{2}\right)\right)}
{(1-u)\left(1+p\left( u+u^{2}\right)\right)}}du,\label{eq:17}
\end{equation}

\begin{figure}[!ht]
\centering
\includegraphics[width=\linewidth]{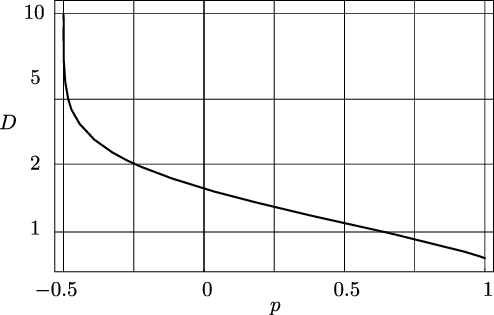}
\caption{Dependence of the deformation factor $D$ on the ``dispersion'' parameter $p$. The graph is presented on a logarithmic scale.}\label{fig3}
\end{figure}

According to the graph presented in Fig.~\ref{fig3}, the strongest deformation is experienced by thermal spots in the negative region of the parameter $-0.5<p\leq0$. Moreover, in this region of the parameter $p$, ``hot'' thermal spots ($\gamma=-1$) move in the eastward direction ($c>0$), while ``cold'' ones ($\gamma=1$) move in the westward direction ($c<0$). In addition, we note that as $p$ approaches its lower limiting value ($p\rightarrow-0.5$), the equatorial semi-axes of the spots elongate, and on the sphere they are bounded by the length of the equator. They experience unbounded growth, as shown in Fig.~\ref{fig3}, only in the equatorial beta-plane approximation.

\section{Self-similar solutions}\label{sec4} 

Another physically important class of solutions possessed by equations~\eqref{eq:1} is that of solutions describing self-similar thermal spots. To study solutions of this kind, by analogy with~\cite{g25}, we first perform a natural parametrization of the model, choosing the natural parameter $\sigma$ as the new independent variable and thereby imposing the constraint
\begin{equation}
X_{\sigma}^{2}+Y_{\sigma}^{2}=1,\label{eq:18}
\end{equation}
and, in addition, we shall assume the cross-frontal buoyancy gradient $\gamma$ to be a parameter depending on time. Although such a dependence is not assumed in the original conservative model, it is reasonable to suppose that for processes of sufficiently slow external (adiabatic) heating or cooling of thermal spots such a parametrization is possible. In particular, one more argument in favor of the nonstationarity of $\gamma$ is that even in the absence of thermal losses the frontal temperature jump of a spot may depend on its area or perimeter.

Under the natural parametrization, the variable $s$, along with the variables $X$ and $Y$, becomes a function depending on $\sigma$ and $t$, and equations~\eqref{eq:1} are first transformed to the form
\begin{gather}
\left(X_{t}s_{\sigma}-s_{t}X_{\sigma}\right)Y=\gamma\partial_{\sigma}
\left[s_{\sigma}\left(s_{\sigma}X_{\sigma\sigma}-s_{\sigma\sigma}X_{\sigma}\right)\right],\label{eq:19}\\
\left(Y_{t}s_{\sigma}-s_{t}Y_{\sigma}\right)Y=\gamma\partial_{\sigma}
\left[s_{\sigma}\left(s_{\sigma}Y_{\sigma\sigma}-s_{\sigma\sigma}Y_{\sigma}\right)\right].\label{eq:20}
\end{gather}
and then the use of the substitution $s=\pm Y$, which turns~\eqref{eq:20} into the identity $0\equiv0$, and of condition \eqref{eq:18}, which linearizes the right-hand side of~\eqref{eq:19}, brings us to the problem
\begin{equation}
\left(X_{t}Y_{\sigma}-Y_{t}X_{\sigma}\right)Y=\pm\gamma X_{\sigma\sigma\sigma },\quad
X_{\sigma}^{2}+Y_{\sigma }^{2}=1,\label{eq:21}
\end{equation}

As is easy to verify, the resulting equations are invariant under translations along the $X$ axis, under the substitutions $X\rightarrow-X$, $Y\rightarrow-Y$, and possess self-similar solutions
\begin{equation}
Y=Ry,\quad X=X_{0}+Rx,\label{eq:22}
\end{equation}
where $X_{0}$ is a translational shift, $R=R(t)$ is a function of time, and the variables $y$ and $x$ depend only on the self-similar parameter $\xi =\sigma /R$ and satisfy the equations
\begin{gather}
y\left(xy^{\prime}-yx^{\prime }\right)=x^{\prime\prime\prime},\quad
x^{\prime 2}+y^{\prime 2}=1,\label{eq:23}\\
R^{3}R_{t}=\pm\gamma.\label{eq:24}
\end{gather}
Here and below a prime denotes differentiation with respect to the self-similarity parameter $\xi$.

Since we are interested in solutions with a closed contour, by virtue of the mirror symmetry with respect to the coordinate axes, for a qualitative analysis it suffices to consider the behavior of the curve in a neighborhood of the turning point located on the positive semi-axis $y$. Assuming that at this point $\xi=0$, $y(0)=y_{0}$, and $x(0)=0$, one can find the expansions in a neighborhood of this turning point
\begin{gather}
x=\xi-\frac{y_{0}^{2}}{3}\xi^{3}+\frac{y_{0}^{4}}{120}\xi^{5}+\cdots,\label{eq:25}\\
y=y_{0}-\frac{y_{0}}{2}\xi^{2}+\frac{y_{0}^{3}}{24}\xi^{4}+\cdots.\label{eq:26}
\end{gather}

\begin{figure}[t!]
\centering{\includegraphics[width=\linewidth]{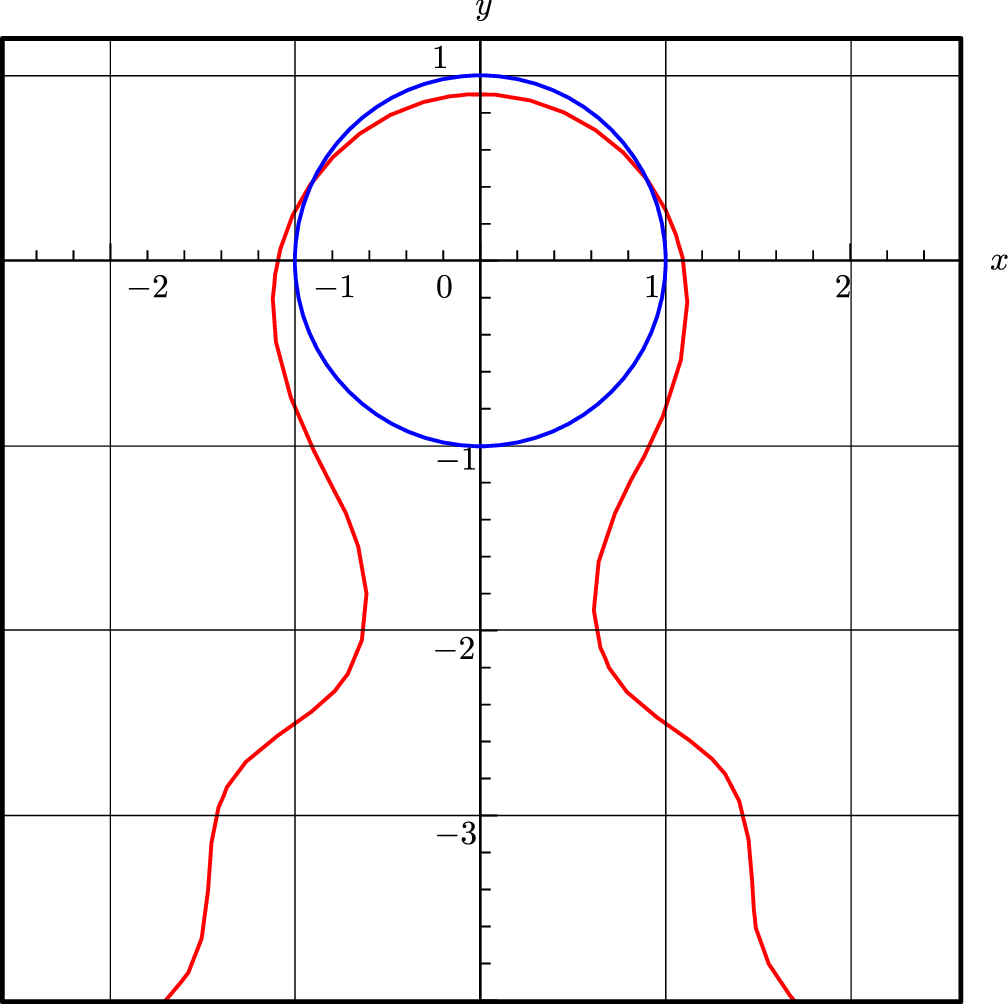}}
\caption{Opening of the contour upon departure from the regime $y_{0}=1$ (light blue curve) to the regime $y_{0}=0.9$ (red curve).}\label{fig4}
\end{figure}

In essence, these expansions say that the curves satisfying the third-order ODE form not a two-parameter but a one-parameter family, depending only on the amplitude value of the curve $y_{0}$. Moreover, one can verify directly that for $y_{0}=1$ the solution represents a circle:
\begin{equation}
y=\cos\frac{\sigma}{R},\quad x=\sin\frac{\sigma}{R},\label{eq:27}
\end{equation}
and is, as numerical analysis shows, the only regular solution in the class of closed curves for equations~\eqref{eq:23}. The loss of closedness upon departure from the regime $y_{0}=1$ is demonstrated in Fig.~\ref{fig4}.

According to~\eqref{eq:24}, depending on the choice of sign for the same $\gamma$, different scenarios may be realized for the temporal behavior of the spot radius. If $\gamma=\const$, then, depending on the sign of the right-hand side, thermal spots may either expand unboundedly according to the law
\begin{equation}
R=R_{0}\left(1+\frac{t}{\tau}\right)^{1/4},\quad\tau=\frac{R_{0}^{4}}{4|\gamma|},\quad t\geq0;\label{eq:28}
\end{equation}
or collapse as
\begin{equation}
R=R_{0}\left(1-\frac{t}{\tau}\right)^{1/4},\label{eq:29}
\end{equation}
degenerating into a point in finite time. Here $R_{0}$ is the initial radius of the spot, and the positive constant $\tau$ is the characteristic time.

As already noted above, the assumption that the cross-frontal gradient $\gamma$ is constant on the contour of thermal spots, which was ``built into'' the original conservative model, in some regimes leads to paradoxical results that may even seem devoid of sound physical meaning.

For example, a certain cognitive dissonance may be caused by the constancy of $\gamma$ under self-similar expansion of a spot in the absence of heat exchange. Within the framework of the model considered here, the simplest and most physically natural way to restore common sense is to use additional constraints, for example, conservation laws. In particular, if one stays within the framework of equation~\eqref{eq:24} and assumes conservation of the total buoyancy of the spot, then an additional condition arises
\begin{equation}
\gamma R^{2}=\gamma_{0}R_{0}^{2},\label{eq:30}
\end{equation}
where $\gamma_{0}$ is the initial buoyancy gradient. The use of this condition in solving equation~\eqref{eq:24} leads to a modification of the result obtained in~\eqref{eq:28}:
\begin{gather}
R=R_{0}\left(1+\frac{t}{\tau}\right)^{1/6},\quad \gamma=\gamma_{0}\left(1+\frac{t}{\tau}\right)^{-1/3},\label{eq:31}\\
\tau=\frac{R_{0}^{4}}{6|\gamma_{0}|},\quad t\geq0.\label{eq:32}
\end{gather}
which gives a more convincing picture of the self-similar behavior of a thermal spot. Namely, almost as in~\eqref{eq:28}, the increase in the spot radius occurs in a power-law manner, but more slowly, and, importantly, is accompanied by a decrease in the cross-frontal gradient, down to its complete disappearance at $t=\infty$.

\section{Discussion and conclusions}\label{sec5} 

The qualitative characteristics of thermal spots as functions of the shape factor $b/a$, which fixes the parameter $p$, and of the sign of the temperature jump $\Delta T$ are presented in Table\,1. In the upper right cell of the table, formulas for determining the speed of the spots are given. As follows from the results obtained, the speed and direction of motion of the spots depend not only on the sign of their thermal contrast with the background flow, but also on the ratio of the semi-axes determining their shape. In particular, if the ratio of the equatorial semi-axis $b$ to the meridional one $a$ is equal to $\pi/2$, the spots are at rest regardless of the sign of the thermal contrast. If the spots are elongated along the equatorial axis to such an extent that $b/a>\pi/2$, then ``cold'' spots move westward and ``hot'' spots move eastward. Under the condition $0.747\leq b/a<\pi/2$, the situation is reversed: ``hot'' spots move westward and ``cold'' spots move eastward.
\begin{table}[t!]
  \centering
\caption{Qualitative characteristics of thermal spots moving along the equator.}
\vspace{10pt}
\begin{tabular}{|c|c|c||c|}
\hline
$b/a>\pi/2$ & $b/a=\pi/2$ & $\pi/2\geq b/a>0.747$ &
\multirow{2}{*}{$p=\dfrac{\gamma}{3}ca^{3},\quad \dfrac{b}{a}=D(p)$}\\
\cline{1-3}
$-0.5<p<0$ & $p=0$ & $0<p\leq1$ &\\
\hline\hline
$c>0$ & \multirow{2}{*}{$c=0$} & $c<0$ & $\Delta T>0,\quad\gamma=-1$\\
\cline{1-1}\cline{3-4}
$c<0$ & & $c>0$ & $\Delta T<0,\quad\gamma=1$\\
\hline
\end{tabular}
\end{table}

\begin{figure}[t!]
\centering{\includegraphics[width=\linewidth]{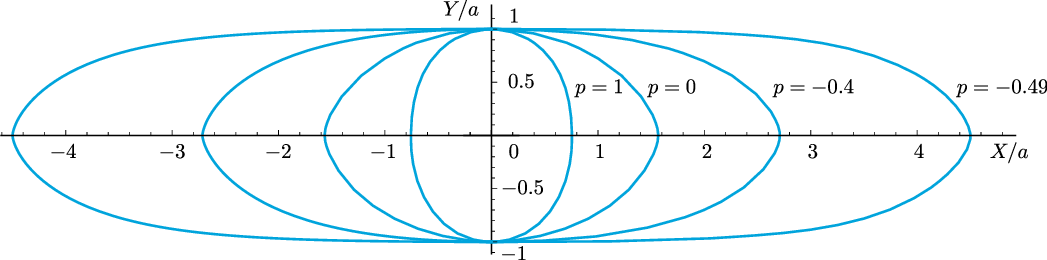}}
\caption{Hierarchy of equatorial thermal spots as functions of the ``dispersion'' parameter $p$.}\label{fig5}
\end{figure}

In conclusion, let us give numerical estimates assuming that the characteristic time scale coincides with the half-period of the Earth's rotation ($T=12\,\text{h}$), and the length scale with the equatorial Rossby deformation radius ($L\approx 200\,\text{km}$). Then, in accordance with the results obtained above, for an eastward-moving elongated ``cold'' spot with $p=0.2$ (which corresponds to $D=b/a=1.44$ and $\gamma=1$) and a meridional semi-axis of $100\,\text{km}$ ($a=0.5$), from the formula $p=\gamma ca^{3}/3$ we find $c=4.8$. In dimensional units this amounts to approximately $22\,\text{m/s}$. A selective hierarchy of thermal spots, which gives a qualitative idea of the change in their shape as the parameter $p$ varies, is presented in Fig.~\ref{fig5}.

As for the self-similar stage of spot evolution, it follows from~\eqref{eq:28} and~\eqref{eq:31} that it is characterized by power-law scaling with exponents $k=1/4$ and $k=1/6$, depending on whether the cross-frontal buoyancy gradient is assumed constant or the total buoyancy of the spot is conserved. In both cases the solution depends on the initial condition only through the amplitude factors $R_{0}$ and $\gamma_{0}$, whereas the exponent itself turns out to be universal. Consequently, on sufficiently long times the self-similar solutions ``forget'' the initial conditions, and the shape of the spot is determined solely by the self-similarity parameter~\cite{kuz04,knnr07}. It is precisely this property that makes them natural candidates for the role of structural elements in determining the low-frequency asymptotics of turbulence spectra at synoptic scales.

\section*{Declaration of Competing Interest}
The author declare that he has no known competing financial interests or personal relationships that could have appeared to influence the work reported in this paper.

\section*{Acknowledgments}
This work was supported by the Russian Science Foundation No. 23-17-00273.

\bibliographystyle{elsarticle-harv}

\begin{thebibliography}{99}

\bibitem[Deem and Zabusky(1978)]{dg78}
G.\,S. Deem and N.\,J. Zabusky,
``Vortex waves: Stationary <<V states>>, interactions, recurrence and breaking'',
Phys. Rev. Lett. \textbf{40}, 859 (1978).
\url{https://doi.org/10.1103/PhysRevLett.40.859}

\bibitem[Pullin(1992)]{p92}
D.\,L. Pullin,
``Contour dynamics methods'',
Annu. Rev. Fluid Mech. \textbf{24}, 9 (1992).
\url{https://doi.org/10.1146/annurev.fl.24.010192.000513}

\bibitem[Zeitlin(2018)]{z18}
V. Zeitlin, 
{\sl Geophysical Fluid Dynamics: Understanding (Almost) Everything with Rotating Shallow Water Models}. Oxford University Press, 2018.
http://dx.doi.org/10.1080/00107514.2018.1515254.

\bibitem[Goncharov(2023)]{g23}
V.\,P. Goncharov,
``Influence of the beta-effect on dynamics of frontal temperature jets'',
Phys. Rev. Fluids \textbf{35}, 066606 (2023).

\bibitem[Goncharov(2025)]{g25}
V.\,P. Goncharov, 
``Dynamics of frontal jets and heat spots in a rotating layer'',
Dynam. Atmos. Oceans \textbf{111}, 101584 (2025)
\url{https://doi.org/10.1016/j.dynatmoce.2025.101584}

\bibitem[Rudenko(2018)]{r18}
O.\,V. Rudenko, 
``<<Exotic>> models of high-intensity wave physics: linearizing equations, exactly solvable problems and non-analytic nonlinearities'', (In Russian)
Izvestiya vysshikh uchebnykh zavedeniy. Prikladnaya Nelineynaya Dinamika \textbf{26}:3, 7 (2018).
\url{https://doi.org/10.18500/0869-6632-2018-26-3-7-34}

\bibitem[Goncharov(2021)]{g21}
V.\,P. Goncharov,
``Dynamics of thin jets generated by temperature fronts'',
Phys. Rev. Fluids \textbf{6}, 103801 (2021).
\url{https://link.aps.org/doi/10.1103/PhysRevFluids.6.103801}...

\bibitem[Kuznetsov(2004)]{kuz04}
E.\,A. Kuznetsov,
``Turbulence spectra generated by singularities'',
JETP Lett. \textbf{80}, 83 (2004).
https://doi.org/10.1134/1.1804214

\bibitem[Kuznetsov et al.(2007)]{knnr07}
E.\,A. Kuznetsov, V. Naulin, A.\,H. Nielsen, and J.\,J. Rasmussen,
``Effects of sharp vorticity gradients in two-dimensional hydrodynamic turbulence'',
Phys.~Fluids \textbf{19}, 105110 (2007).
\url{https://doi.org/10.1063/1.2793150}

\end{thebibliography}

\end{document}